\documentclass[11pt,a4paper]{article}
\usepackage{jheppub_kim}

\usepackage{pdflscape}
\usepackage{amsmath}
\usepackage{amssymb}
\usepackage{dcolumn}
\usepackage{bm}
\usepackage{color}
\usepackage{epsfig}
\usepackage{amsfonts}
\usepackage{graphicx}
\usepackage{subfigure}
\usepackage{dcolumn}

\newcommand{\be}{\begin{equation}}
\newcommand{\ee}{\end{equation}}
\newcommand{\bea}{\begin{eqnarray}}
\newcommand{\eea}{\end{eqnarray}}

\def\be{\begin{equation}}
\def\ee{\end{equation}}
\def\bea{\begin{eqnarray}}
\def\eea{\end{eqnarray}}

\begin{document}

\title{Holographic Schwinger effect with a deformed AdS background}
\author[a]{J. Sadeghi}
\author[b]{B. Pourhassan}
\author[a]{S. Tahery}
\author[a]{F. Razavi}

\affiliation[a]{Sciences Faculty, Department of Physics, University of Mazandaran, 47416-95447, Babolsar, Iran}
\affiliation[b]{School of Physics, Damghan University, 36716-41167, Damghan, Iran}

\emailAdd{pouriya@ipm.ir}
\emailAdd{b.pourhassan@du.ac.ir}
\emailAdd{s.tahery@stu.umz.ac.ir}
\emailAdd{f.razavi@stu.umz.ac.ir}

\abstract{In this paper, we consider a deformed AdS background and study effect of deformation parameter on the pair production rate of the Schwinger effect. The electrostatic potential is important for the pair production in the holographic Schwinger effect. In this paper, we analyze the electrostatic potential in a deformed AdS background and investigate the effect of deformation parameter which may be useful to test of AdS/QCD. In the case of zero temperature we find that larger value of the deformation parameter leads to a smaller value of separation length of the test particles on the probe. Also we find a finite maximum of separation length in presence of modification parameter.}

\keywords{AdS/CFT; Schwinger effect; QCD; QED.}

\maketitle

\section{Introduction}
An interesting non-perturbative phenomena, which is production of pair in an external electric field in quantum electrodynamics
 (QED) is called Schwinger effect \cite{jss}, where due to a presence of a strong electric-field, virtual electron-positron pair become
real particles. This production rate can be calculated using  imaginary part of Euler-Heisenberg Lagrangian \cite{wh}. It is not restricted to QED and can be used for any charged particles in a strong external field such as an electromagnetic field. Furthermore, it can lead to the creation of a neutral pair of higher dimensional objects such as string and D-branes. Potential analysis is an important stage to study Schwinger effect. In the context of QED, the potential analysis estimated by the static potential including Coulomb interaction between the particles. In addition, an energy $Ex$, where $x$ is a separating distance of virtual pairs and $E$ is an external electric field, should be taken into account. When the virtual pair get a greater energy than the rest energy from an external electric field, they become real. So, for the creation of a real pair which corresponds to the vacuum decay rate, the external field should reach to a critical value, where the vacuum becomes totally unstable. Value of the critical electric field is beyond in which the effective Euler-Heisenberg action obtains an imaginary part. In the other words, when the electric field is small, the potential barrier is present and the pair production is described as a tunneling process. the potential barrier decreases as the electric field becomes greater and at a critical value of $E$, the barrier vanishes and the production rate is catastrophic and is not exponentially suppressed. The critical value of the electric field is regarded as the critical behavior of UV completion of the string in string theory point of view \cite{esf,cbm}. On the other hand, in QCD point of view, the vacuum decay rate for Schwinger effect is related to the quark antiquark pair creation in the presence of a strong field.\\
Recently, AdS/CFT correspondence used to study Schwinger effect. AdS/CFT correspondence, which is relation between a $d$-dimensional conformal field theory (CFT) and a $(d+1)$-dimensional string theory in anti-de Sitter (AdS) space, is a powerful mathematical tools to investigate about strongly correlated systems \cite{P1, P2, P3, oas}. The role of an extra dimension in the AdS side may be understood using the energy scale of the CFT side on the boundary. AdS/CFT correspondence already used with success to study properties of quark-gluon plasma (QGP) \cite{P4, P5, B1, B2, B3, B4}. It leads to an analytic semi-classical model for strongly coupled QCD. It has scale invariance, dimensional counting at short distances and color confinement at large distances. This theory describes the phenomenology of hadronic properties and demonstrates their ability to incorporate such essential properties of QCD as a confinement and a chiral symmetry breaking. From the AdS/CFT point of view the $AdS_5$ plays an important role in describing QCD phenomena so it called AdS/QCD \cite{AQ1,AQ2,AQ3,AQ4,B5}.\\
Since one may consider a quark-antiquark pair creation as well as the electron-positron pair, many works have been done about Schwinger effect in a holographic setup related to quark-antiquark pair. The creation rate of the quark pair in $N=4$ SYM theory was obtained in \cite{stp} and based on them, the holographic  Schwinger effect were calculated in various systems \cite{rhw,coce,hds,secp,uah,ppr,hdse,npp,hsed,nre,scen}. Also the vacuum decay rate is regarded as the creation rate of the quark-antiquark in $N=2$ SQCD \cite{vie}. In the Ref. \cite{pah} electrostatic potentials in the holographic Schwinger effect has been analyzed for the finite-temperature and temperature-dependent critical-field cases to find  agreement with the DBI result. In the Ref. \cite{hse} tunneling pair creation of W-Bosons by an external electric field on the Coulomb branch of N=4 supersymmetric Yang-Mills theory has been studied and found that the pair creation formula has an upper critical electric field beyond which the process is no longer exponentially suppressed. We obtain such result in our work with the deformed AdS background. As we know from the SYM theory, the supersymmetry and conformal symmetry are broken at finite temperature which made some problem to test predictions based on the AdS/CFT correspondence. The best way is the construction of dual gravity where the conformal symmetry
is not broken, and it is possible with deformation of the AdS space \cite{CLSS}. In that case there are several way to deform AdS space in agreement with
experimental data and lattice QCD results \cite{Ex1,Ex2,cor,hqp}. Deformation parameter of AdS space may related to the electromagnetic field, on the other hand electromagnetic field may related to non-commutative space \cite{Xing} so it is completely reasonable that any modification in the space related to the electrostatic potential. In the recent work \cite{GI} the supersymmetric Yang-Mills theory considered to study instability (related to the Schwinger effect)
caused by the external electric field through the imaginary part of the action of
the D7 probe brane embedded in the background of type IIB theory. It is clear that there is almost 30 percent deviation from both the string world-sheet result and the DBI result. In \cite{pah} this deviation having been reconsidered from the viewpoint of the potential analysis. Therefore, any modification of background may affect experimental results, hence in this paper we would like to consider deformed AdS metric to study the effect of deformation parameter in the potential and critical field. It will be useful to study holographic Schwinger effect and test of AdS/QCD. In that case one can start from an effective field theory somehow motivated by string theory and try to fit with QCD. In the Ref. \cite{cor}, from QCD analysis of two current correlators a quadratic correction is exist. On the other hand from AdS/QCD correspondence, it has been tried to find a string description of strong interactions to address the issue of the quadratic correction within the simplified model \cite{hsc,scv} and finally the basic framework was set by considering a metric background which behaves asymptotically as $AdS_5\times X$, where $X$ is some five dimensional compact space. By giving the metric background in \cite{hqp}, they attempt to calculate the heavy quark potential. Also, the results of modeling the temperature dependence of the spatial string tension and thermal phase transition in a five-dimensional framework have been studied by the Ref. \cite{tsst}. The effects of deformation parameter on thermal width of moving quarkonia in plasma has been studied by the recent work \cite{tec}.\\
This paper is organized as follows. In the next section we give brief review of Schwinger effect, then in sections 3 and 4 give potential analysis of zero and finite temperature case respectively. Finally in section 5 we give conclusion and summary of results together outlook of future works.

\section{Schwinger effect}
The pair production rate in an external electric field $E$ with weak coupling calculated by Schwinger as follow \cite{jss},
\begin{equation} \label{eq:rate1}
\Gamma\sim e^{-\frac{\pi m^{2}}{eE}}.
\end{equation}
In this case, there is no critical field trivially. This extended to the arbitrary coupling as \cite{AAM},
\begin{equation} \label{eq:rate2}
\Gamma\sim e^{-\frac{\pi m^{2}}{eE}+\frac{e^{2}}{4}}.
\end{equation}
The production rate of the fundamental particle becomes related to the critical field as \cite{hse},
\begin{equation}\label{eq:production rate}
\Gamma\sim exp\left[ -\frac{\sqrt{\lambda}}{2}\left( \sqrt{\frac{E_{cr}}{E}}-\sqrt{\frac{E}{E_{cr}}}\right)^2\right]  ,\quad\quad E_{cr}=\frac{2\pi m^2}{\sqrt{\lambda}},
\end{equation}
where $m$ is mass and $\lambda$ is 't Hooft coupling. One can check that the equation (\ref{eq:production rate}) is in agreement with DBI result. On the other hand, from the AdS/CFT point of view and using the Coulomb potential, the critical field has been obtained as \cite{msh,wll},
\begin{equation}\label{eq:critical000}
E_{cr}\sim 0.70 \frac{2\pi m^2}{\sqrt{\lambda}}.
\end{equation}
In order to study Schwinger effect using the AdS/CFT correspondence, the setup is to put a probe D3-brane at an intermediate position $z=z_0$ rather than close to the boundary \cite{hse}. Hence, the mass becomes finite and we have,
\begin{equation} \label{eq:eta zero}
g_{ab}=diag(-\frac{R^2}{z^2}, \frac{R^2}{z^2}),
\end{equation}
where $g_{ab}$ is the induced metric on the string world-sheet.  Therefore, the mass is given by,
\begin{equation} \label{eq:m zero}
m=T_F\int_{z_0}^{\infty} dz \sqrt{-det g_{ab}},
\end{equation}
where $T_F=\frac{1}{2\pi\alpha'}$ is the string tension.\\
In the next sections we will study Schwinger effect in a deformed AdS background by discuss about the potential.

\section{Deformed AdS at zero temperature}
First of all, we consider deformed AdS background at zero temperature and investigate the effect of deformation parameter on the sum of Coulomb potential (CP) and static energy (SE) denoted by $V_{CP+SE}$. We do that by calculation of mass given by deformed AdS metric. In that case, using the critical electric field we will find total potential.\\
We begin with the deformed AdS metric given by \cite{hqp},
\begin{equation} \label{metric zero}
ds^2=\frac{R^2}{z^2} h(z) (\eta_{\mu\nu}\Sigma_{i=0}^3 dx_i ^2+dz^2)+R^2 d\Omega_{5}^2,\quad \quad h(z)=e^{\frac{cz^2}{2}},
\end{equation}
where $R$ is radius of space which is related to the sloe parameter and coupling via, $R^2=\alpha' \sqrt{\lambda}$, with $\alpha'=l_s^2$ where $l_s$ is the string scale. Moreover $d\Omega_5^2$ is metric of five-dimensional sphere. We should note that the signature of $\eta_{\mu\nu}$ will use to evaluate the potential. In the former case, Euclidean signature $\eta_{\mu\nu}=(1,1,1,1)$ is used, while in the DBI action, the Lorentzian signature $\eta_{\mu\nu}=(-1,1,1,1)$ is considered to escape from transformation of the electric field to the magnetic field in the case of Wick-rotation.\\
The quark antiquark potential can obtain using the expectation value of the Wilson loop. The loop is corresponds to a trajectory of test particles with infinite heavy mass, and the expectation value is corresponds to the area of a string world-sheet attached to the Wilson loop \cite{msh,wll}.\\
In order to study Schwinger effect, we should extend relation (\ref{eq:eta zero}) to the case including deformation parameter, so we have
\begin{equation} \label{eq:eta zero-c}
g_{ab}=diag(-\frac{R^2}{z^2}e^{\frac{cz^2}{2}}, \frac{R^2}{z^2}e^{\frac{cz^2}{2}}),
\end{equation}
Therefore, using the equation (\ref{eq:m zero}) one can obtain the following mass,
\begin{equation} \label{eq:m zero-c}
m= T_F \int_{z_0}^{\infty} dz \frac{R^2}{z^2} e^{\frac{cz^2}{2}},
\end{equation}
It is clear that the mass increases with increasing $c$, but it should notice that this expression does not make difficulties with divergent mass, because the deformation parameter has an upper limit  phenomenologically \cite{cor,hqp,tsst} and then the mass does not diverge even if one considers heavy particles.
At the next step, we will compute the $V_{CP+SE}$ and discuss the critical electric field.\\

Now, we use AdS/CFT to study Coulomb potential plus static energy. In that case we should consider the area of rectangular Wilson loop on the probe D3-brane, and evaluate the classical action of a string attached to the probe D3-brane. We follow the circular Wilson-loop analysis of the Ref. \cite{hse}, but with the deformed AdS background. The Nambo-Goto string action is given by,

\begin{eqnarray} \label{eq:NG}
S&=&T_F \int {d\tau d \sigma \mathcal{L}}\nonumber\\
&=&T_F  \int {d\tau d \sigma \sqrt{det G_{ab}}},
\end{eqnarray}
where
\begin{equation}\label{G}
G_{ab}\equiv \frac{\partial x^\mu}{\partial \sigma^{a}} \frac{\partial x^\nu}{\partial \sigma^{b}} g_{\mu\nu},
\end{equation}
is the induced metric and $\sigma^{a}=(\tau, \sigma)$ are world-sheet coordinates. It is useful to choose the static gauge, $x^{0}=\tau$, and $x^{1}=\sigma$. So, the radial direction $z(\sigma)$ depends only on $\sigma$ in classical solution.
Therefore, the Lagrangian is,
\begin{equation} \label{eq:L zero}
\mathcal{L}=\frac{R^2}{z^2} e^{\frac{cz^2}{2}} \sqrt{1+(\frac{dz}{d\sigma})^2}.
\end{equation}
So, from the equation of motion, one can find,
\begin{equation} \label{eq: eqmo}
\frac{\partial \mathcal{L}}{\partial (\partial _{\sigma} z)} \partial_{\sigma} z -\mathcal{L}=C,
\end{equation}
where $C$ is an arbitrary constant, and this yields to the following relation,
\begin{equation} \label{eq:cte zero}
\frac{R^2}{z^2}\frac{ e^{\frac{cz^2}{2}}}{\sqrt{1+(\frac{dz}{d\sigma})^2}}=C,
\end{equation}
again, $C$ is an arbitrary constant. The important boundary condition at $\sigma=0$, imposes,
\begin{equation} \label{eq:bouncon zero}
\frac{dz}{d\sigma}=0, \quad \quad z=z_{\ast},
\end{equation}
where $z_{\ast}$ is the turning point, which means the deepest position of the string in the bulk. Therefore, we yield to the following differential equation,
\begin{equation} \label{eq:difeq zero}
\frac{dz}{d\sigma}=\sqrt{\frac{z_\ast ^4}{z^4} e^{c(z^2-z_\ast ^2)}-1} .
\end{equation}
By using  the change of variables $y=\frac{z_{\ast}}{z}$ and $a=\frac{z_0}{z_{\ast}}$, in the equation (\ref{eq:difeq zero}) one can obtain the separation length of the test particles on the probe brane as,
\begin{eqnarray}\label{eq:x zero}
x&=&2z_{\ast}\int_{1}^{\frac{1}{a}} \frac{dy}{y^2\sqrt{y^4 e^{cz_{\ast}^2 (\frac{1}{y^2}-1)}-1}}\nonumber\\
&=&2\frac{z_{0}}{a} \int_{1}^{\frac{1}{a}} \frac{dy}{y^2\sqrt{y^4 e^{c\frac{z_{0}^2}{a^2} (\frac{1}{y^2}-1)}-1}}.
\end{eqnarray}
In the case of $c=0$, the separation length obtained in terms of Gamma and hypergeometric functions \cite{pah}. In presence of modification parameter, it is hard to obtain analytical expression of the above integral, so we give numerical analysis and draw $x$.

\begin{figure}[h!]
\begin{center}$
\begin{array}{cccc}
\includegraphics[width=100 mm]{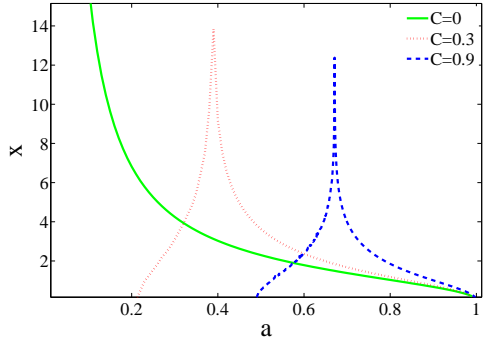}
\end{array}$
\end{center}
\caption{The distance $x$ against $a$ for different values of $c$.}
\label{fig:xatzero}
\end{figure}

Fig. \ref{fig:xatzero} illustrates the distance $x$ with respect to $a$ at zero temperature overall with contribution of the deformation parameter. The parameter $a$ shows the position of the tip of the string. In the case of $c=0$, we can see $x$ decreases with increasing $a$, the branch with smaller values of $a$ means the region close to the horizon and describes another configuration of the string world-sheet, a pair of straight lines as discussed in \cite{wpl}. So, the larger values of $a$ has been taken for the present analysis. When the deformation parameter contributes, the behavior of $x$ changes significantly. On one hand, the region close to the horizon, having been mentioned above, corresponds to the more values of $a$ in comparison with the case $c=0$. Therefore with increasing deformation parameter, the change in the configuration of string world-sheet takes place further from the horizon. On the other hand, with contribution of $c$, the distance $x$ increases with increasing $a$ at the first step till it attains a maximum value, then it descends. In any deformed AdS space, there is a unique position where the separation length of the test particles is maximum. The maximum value of the $a$ decreases with increasing $c$. It can be interpreted as a larger value of the deformation parameter, leads to a smaller value of separation length of the test particles on the probe. In addition, there is a degeneracy between the distance $x$ and the the auxiliary parameter $a$ at nonzero values of $c$, while at $c=0$ a single value of $x$ corresponds to a unique value of $a$. Another important result is presence of maximum  value of separation length, which has finite value for $c\neq0$, while infinite value for $c=0$ corresponds to $a=0$. Therefore, it seems presence of modification parameter is crucial at zero temperature to avoid divergency of separation length.\\
Now, the sum of the Coulomb potential and static energy will be obtained  through (\ref{eq:L zero}), (\ref {eq:difeq zero}) and (\ref{eq:x zero}) as,
\begin{eqnarray}\label{eq:vcp zero}
V_{CP+SE}&=&2T_F\int_{0}^{\frac{x}{2}} dx \mathcal{L}\nonumber\\
&=&2T_F \frac{R^2}{z_0} a \int_{1}^{\frac{1}{a}} \frac{y^2 e^{c\frac{z_{0}^2}{a^2} (\frac{1}{y^2}-\frac{1}{2})}dy}{\sqrt{y^4 e^{c\frac{z_{0}^2}{a^2} (\frac{1}{y^2}-1)}-1}}.
\end{eqnarray}

\begin{figure}[h!]
\begin{center}$
\begin{array}{cccc}
\includegraphics[width=100 mm]{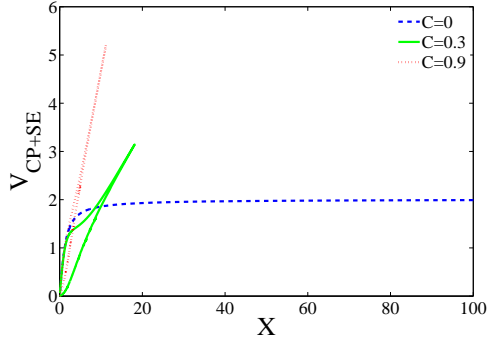}
\end{array}$
\end{center}
\caption{$V_{CP+SE}$ versus $x$ at zero temperature with $T_F\frac{R^2}{z_0}=z_0=1$.}
\label{fig:vxtzero}
\end{figure}

In the case of $c=0$ one can obtain analytical expression of integral in the equation (\ref{eq:vcp zero}) and find $V_{CP+SE}$ in terms of Gamma and hypergeometric functions \cite{pah}. In presence of deformation parameter ($c\neq0$) we need numerical analysis to find behavior of potential.
In that case, Fig. \ref{fig:vxtzero} shows $V_{CP+SE}$ versus $x$. At $c=0$, potential behaves exponentially at short distances and then it is changed to a monotonous manner. In presence of the deformation parameter, there is no difference between variable graphs with various $c$ at very short distances. It means that deformed AdS behaves as ordinary AdS at short distance. On the other hand, the presence of deformation parameter leads to the existence of a maximum value of the potential where an increase in value of $c$ leads to a greater maximum of $V_{CP+SE}$ in a shorter range of distance  while there is no electric field yet, but the sum of the static and coulomb energy of the test particles in a deformed AdS have different values at different positions (contrary to $c=0$ case). Generally, the effect of the deformation parameter is increasing maximum of the potential. As before we see maximum distance in presence of modification parameter.\\
Before studying the total potential, it should be mentioned that electric field can be used as a fraction of critical electric field $E_{cr}$ in our following calculations, to do this, we need to find an expression for $E_{cr}$ according to the metric background of deformed AdS. According to the Ref. \cite{eih}, one can see that when there is no magnetic field, the critical electric field is given by,
 \begin{equation}\label{eq:Ecr}
E_{cr}=T_F\sqrt{-g_{00}g_{11}}\vert_{IR},
\end{equation}
then the result in zero temperature case is,
\begin{equation}\label{eq:Ecr zero}
E_{cr}=T_F\frac{R^2}{z_0^2} e^{\frac{cz_0^2}{2}}.
\end{equation}
Now, we define a dimensionless value $\alpha$ which depends on $E_{cr}$ when the deformation parameter has exact value of zero as,
\begin{equation}\label{eq:alpha}
\alpha=\frac{E}{E_{cr}\vert_{c=0}}, \quad\quad E_{cr}\vert_{c=0}=T_F\frac{R^2}{z_0^2}.
\end{equation}
At the next step, the electrostatic potential associated with electric field given by (\ref{eq:x zero}), (\ref{eq:vcp zero}) and (\ref{eq:alpha}) as \cite{pah},
\begin{eqnarray}\label{eq:vtot zero}
V_{tot}&=&V_{CP+SE}-Ex\nonumber\\
&=&2T_F\frac{R^2}{z_0}\int_{1}^{\frac{1}{a}} \frac{ay^4e^{c\frac{z_{0}^2}{a^2} (\frac{1}{y^2}-\frac{1}{2})}-\frac{\alpha}{a}}{y^2 \sqrt{y^4 e^{c\frac{z_{0}^2}{a^2} (\frac{1}{y^2}-1)}-1}} dy.
\end{eqnarray}

\begin{figure}[h!]
\begin{center}$
\begin{array}{cccc}
\includegraphics[width=100 mm]{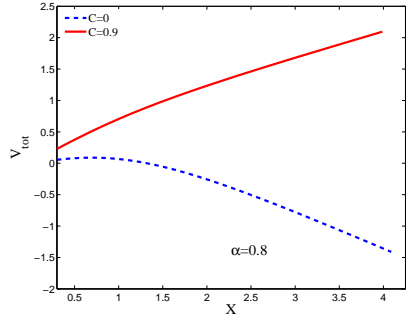}
\end{array}$
\end{center}
\caption{Total potential at zero temperature for $\alpha=0.8$ with $T_F\frac{R^2}{z_0}=z_0=1$.}
\label{fig:vtot1-1}
\end{figure}

\begin{figure}[h!]
\begin{center}$
\begin{array}{cccc}
\includegraphics[width=100 mm]{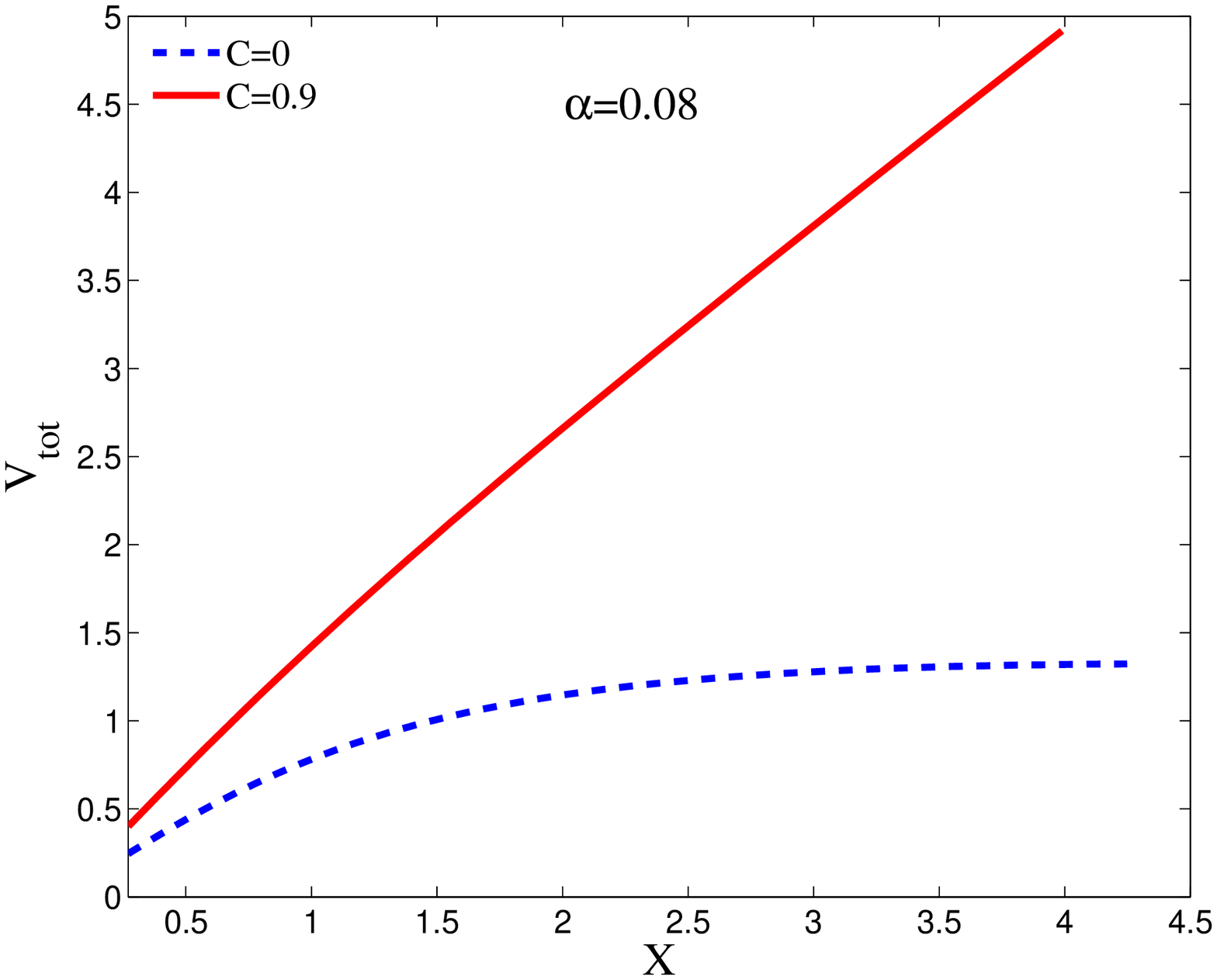}
\end{array}$
\end{center}
\caption{Total potential at zero temperature for $\alpha=0.08$ with $T_F\frac{R^2}{z_0}=z_0=1$.}
\label{fig:vtot1-2}
\end{figure}

\begin{figure}[h!]
\begin{center}$
\begin{array}{cccc}
\includegraphics[width=100 mm]{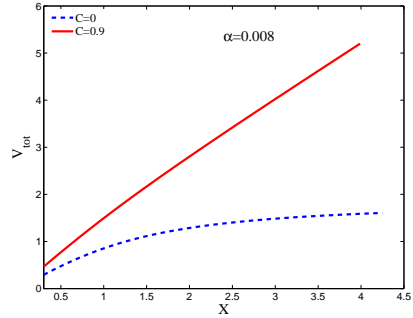}
\end{array}$
\end{center}
\caption{Total potential at zero temperature for $\alpha=0.008$ with $T_F\frac{R^2}{z_0}=z_0=1$.}
\label{fig:vtot1-3}
\end{figure}

\begin{figure}[h!]
\begin{center}$
\begin{array}{cccc}
\includegraphics[width=100 mm]{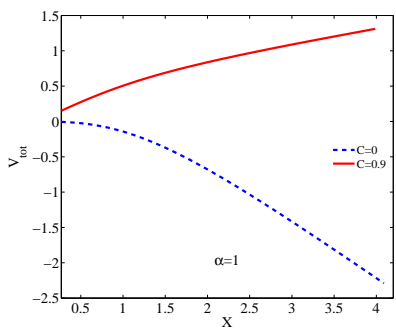}
\end{array}$
\end{center}
\caption{Total potential at zero temperature for $\alpha=1$ with $T_F\frac{R^2}{z_0}=z_0=1$.}
\label{fig:vtot2}
\end{figure}

Fig. \ref{fig:vtot1-1},\ref{fig:vtot1-2},\ref{fig:vtot1-3} and Fig. \ref{fig:vtot2} indicate the behavior of $V_{tot}$ with respect to $x$, where one can see the effects of deformation parameter is  increasing  value of potential at a specified $x$. Furthermore, potential ascends with larger gradient when $c$ has more values. $\alpha=1$ has been defined as follows, electric field has it's critical value and deformation parameter is equal to zero. Therefore, the potential barrier at $c=0$ vanishes for $\alpha\geq 1$. It is interesting that for nonzero values of $c$ the potential barrier does not vanish in this manner. We can compare four potential barriers. The potential barrier vanishes for  $\alpha=1$ at zero value of the deformation parameter. So if electric field has it's critical value at AdS ($c=0$), the pair production process or instability of the vacuum  changes phenomenologically after the contribution of deformation parameter, since there is  tunneling process at any deformed AdS.\\
As the electric field is smaller than it's  critical value, there is a potential barrier which decreases with increasing electric field and increases with increasing deformation parameter. It can be interpreted as the opposed behavior of pair production process in presence of electric field and deformation parameter. Surprisingly, we find that  deformation of AdS metric plays role of decreasing electric field at zero temperature.

\section{Deformed AdS at finite temperature}
In this section we will discuss the effects of deformation parameter at finite  temperature. In order to study potential and separation length we use the following deformed AdS metric \cite{tsst},
\begin{equation}\label{eq:metric finite t}
ds^2=\frac{R^2}{z^2} h(z)  (-f(z) dt^2+\Sigma_{i=1}^{3}dx_i^2+\frac{1}{f(z)} dz^2)+R^2 d\Omega_5^2,\quad\quad h(z)=e^{\frac{cz^2}{2}},
\end{equation}
where
\begin{equation}\label{eq:metric function finite t}
f(z)=1-(\frac{z}{z_h})^4,\quad\quad h(z)=e^{\frac{cz^2}{2}}.
\end{equation}
The horizon is located at $z=z_h$ where $z_h<z_{\ast}<z_0 $, and the temperature of the black hole written as, $T=\frac{1}{\pi z_h}$, so zero temperature limit $z_{h}\rightarrow\infty$ and $f(z)\rightarrow 1$ is discussed by previous section.\\
The DBI action of a probe D3-brane on the deformed background (\ref{eq:metric finite t}) considered as a classical action including an electric field written as  follows,
\begin{eqnarray}\label{eq:action finite t}
S&=&-T_{D3}\int d^4x \sqrt{-det(g_{\mu\nu}+\mathcal{F}_{\mu\nu})}\nonumber\\
&=&-T_{D3} \frac{R^4}{z_0^4} e^{cz_0^2} \sqrt{1-(\frac{z_0}{z_h})^4}\int d^4x \sqrt{1-\frac{(2\pi \alpha_{r})^2 z_0^4}{R^4e^{cz_0^2 (1-(\frac{z_0}{z_h})^4)}} E^2},
\end{eqnarray}
where the tension of D3-brane is given by \cite{cor},
\begin{equation}\label{eq:tension}
T_{D3}=\frac{1}{g_s (2\pi)^3 \alpha_{r}^2}\quad\quad \alpha_{r}=\alpha'\frac{R^2}{z^2} h^{-1}(z).
\end{equation}
It is clear that when the value of the electric field becomes more than the critical electric field, the DBI action is not well-defined. In another words, the vacuum becomes totally unstable as we explained before. The value of $E_{cr}$ at the finite temperature is given by,
\begin{equation}\label{eq:Ecr finite t}
E_{cr}=T_F \frac{R^2}{z_0^2} e^{\frac{cz_0^2}{2}}\sqrt{1-(\frac{z_0}{z_h})^4},
\end{equation}
which depends on deformation parameter in addition to temperature.\\
At the next step, we derive an expression for the mass of the fundamental matter to see the effect of deformation parameter on mass addition to critical electric field. The induced metric of the string world-sheet is given by the equation (\ref{eq:eta zero-c}),
which leads to,
\begin{equation}\label{eq:mass finite t}
m=T_F\int_{z_0}^{z_h} dz \frac {R^2}{z^2}e^{\frac{cz^2}{2}},
\end{equation}
which  depends on temperature and deformation parameter. The only differences with the equation (\ref{eq:m zero-c}) is integration limits. On the other hand, one can use the following relation,
\begin{equation}\label{eq:mt}
m(T)=m(T=0)+\Delta m(T).
\end{equation}
Therefore with (\ref{eq:m zero}), (\ref{eq:mass finite t}) and (\ref{eq:mt}), the following relation should be satisfied,
\begin{eqnarray}\label{eq:delta m}
\Delta m(T)&=&T_F\int_{z_0}^{z_h} dz \frac{R^2}{z^2} e^{\frac{cz^2}{2}}-T_F \int_{z_0}^{\infty} dz \frac{R^2}{z^2} e^{\frac{cz^2}{2}}\nonumber\\
&=&-T_F \int_{z_h}^{\infty} dz \frac{R^2}{z^2} e^{\frac{cz^2}{2}}.
\end{eqnarray}
Proceeding by the following definitions,
\begin{equation}\label{eq:y,a,b}
y=\frac{z_{\ast}}{z},\quad a=\frac{z_0}{z_{\ast}},\quad b=\frac{z_0}{z_h},
\end{equation}
(\ref{eq:Ecr finite t}) and (\ref{eq:delta m}) can be written as,
\begin{equation}\label{eq:Ecr finite t b}
E_{cr}=T_F \frac{R^2}{z_0^2} e^{\frac{cz_0^2}{2}}\sqrt{1-b^4},
\end{equation}
and,
\begin{equation}\label{eq:delta m b}
\Delta m=aT_F\frac{R^2}{z_0}\int_{\frac{1}{b}}^{\infty} dy e^{\frac{cz_0^2}{2a^2y^2}}.
\end{equation}
So, in the presence of deformation parameter, the behavior of  the critical electric field can be studied with respect to $\Delta m$. Here, we can see new parameter $b$ which vanishes at zero temperature ($z_{h}\rightarrow\infty$).\\

\begin{figure}[h!]
\begin{center}$
\begin{array}{cccc}
\includegraphics[width=100 mm]{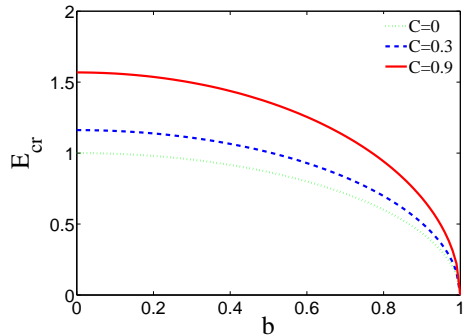}
\end{array}$
\end{center}
\caption{Critical electric field as a function of $b$ with $T_F\frac{R^2}{z_0}=z_0=1$.}
\label{fig:Eb}
\end{figure}

We can see that, both  temperature  and deformation parameter affects critical electric field. Fig.~\ref{fig:Eb} shows behavior of $E_{cr}$ against $b$ at some values of $c$. It has been shown that from zero temperature to finite temperature, with increasing $c$ and $b$, the value of critical electric field increases and decreases respectively. Moreover, the differences between graghs with various deformation parameters are diminished till $b$ reaches to unit. Then all graphs with different $c$ are coincident at large enough temperature. At lower temperatures, more deformation of AdS leads to the more value of critical electric field. Therefore, the pair production process, where vacuum becomes totally unstable, needs more strong electric field.

\begin{figure}[h!]
\begin{center}$
\begin{array}{cccc}
\includegraphics[width=100 mm]{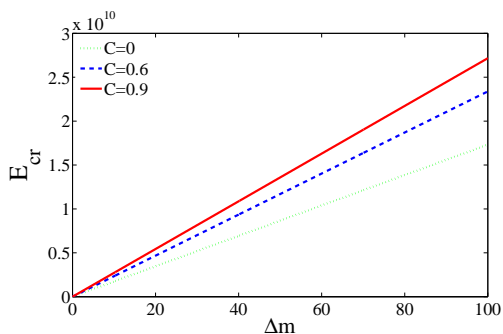}
\end{array}$
\end{center}
\caption{Critical electric field as a function of $\Delta m$ with $T_F\frac{R^2}{z_0}=z_0=1$ and $b=0.5$.}
\label{fig:Edeltam}
\end{figure}

The critical electric field increases linearly with respect to $\Delta m$ as illustrated by the Fig.~\ref{fig:Edeltam}. The affection of temperature and deformation parameter have been mentioned in (\ref{eq:Ecr finite t b}) and (\ref{eq:delta m b}), but it is interesting that the presence of deformation parameter does not change the linear behavior of graphs and just increases gradient of them. On the other hand, for a given particles mass and with a deformed AdS background, the electric field which leads to a pair production is higher than for an AdS with $c=0$. It is worth noting a larger deformation parameter leads to the contribution of a larger critical electric field in the pair production process.\\
Similar to the zero temperature case, the analysis of the potential will be proceeded. The induced metric has been changed, but the approach is as the previous section. According to the equation (\ref{eq:NG}) the Lagrangian can be written as,
\begin{equation} \label{eq:L T}
\mathcal{L}=\frac{R^2}{z^2} e^{\frac{cz^2}{2}} \sqrt{(1-\frac{z^4}{z_h^4})+(\frac{dz}{d\sigma})^2}.
\end{equation}
By using the equation of motion, one can find,
\begin{equation} \label{eq:cte t}
\frac{R^2}{z^2}\frac{ e^{\frac{cz^2}{2}}(1-\frac{z^4}{z_h^4})}{\sqrt{(1-\frac{z^4}{z_h^4})+(\frac{dz}{d\sigma})^2}}=C,
\end{equation}
which yields to the following differential equation,
\begin{equation} \label{eq:difeq t}
\frac{dz}{d\sigma}=\sqrt{\frac{z_\ast ^4}{z^4} e^{c(z^2-z_\ast ^2)}\frac{(1-\frac{z^4}{z_h^4})^2}{(1-\frac{z_{\ast}^4}{z_h^4})}-(1-\frac{z^4}{z_h^4})}.
\end{equation}
The distance $x$ between the particles is obtained from integration of (\ref{eq:difeq t}) as follow,
\begin{equation} \label{eq:x t}
x=2\frac{z_{0}}{a}\sqrt{1-\frac{b^4}{a^4}} \int_{1}^{\frac{1}{a}} \frac{dy}{\sqrt{(y^4-\frac{b^4}{a^4})^2 e^{c\frac{z_{0}^2}{a^2} (\frac{1}{y^2}-1)}-(y^4-\frac{b^4}{a^4})(1-\frac{b^4}{a^4})}},
\end{equation}
where we used definitions of (\ref{eq:y,a,b}).

\begin{figure}[h!]
\begin{center}$
\begin{array}{cccc}
\includegraphics[width=100 mm]{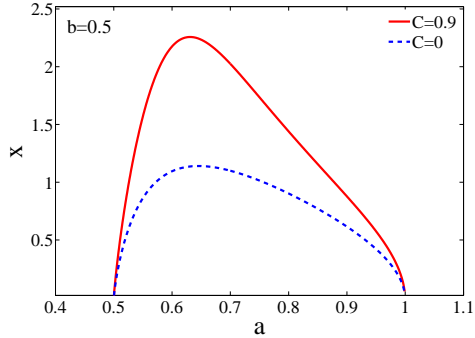}
\end{array}$
\end{center}
\caption{The distance $x$ versus $a$ for nonzero value of $b$ (finite temperature).}
\label{fig:xab}
\end{figure}

Fig. \ref{fig:xab} represents the behavior of the distance $x$ with respect to $a$ for $b=0.5$. It is comparable with the Fig. \ref{fig:xatzero} for $b=0$ and show effect of finite temperature. So, it can be interpreted as thermal case, the behavior of $x$ is completely different and it has a maximum value. In presence of the deformation parameter, the value of maximum increases, or the distance of quark and antiquark  increases in the presence of deformation parameter. On the other hand, at zero temperature the contribution of nonzero deformation parameter leads to a degeneracy between $x$ and $a$, while at finite temperature there is degeneracy at any value of $c$. So here, nonzero value of $c$ or $T$ have similar results phenomenologically.\\
Now, the sum of the Coulomb potential and static energy will be obtained  from (\ref{eq:L T}), (\ref{eq:difeq t}) and (\ref{eq:x t}) as,
\begin{equation} \label{eq:v t}
V_{CP+SE}=2aT_F\frac{R^2}{z_0}\int_{1}^{\frac{1}{a}} \frac{e^{\frac{cz_0^2}{2a^2y^2}}}{\sqrt{1-e^{-c\frac{z_0^2}{a^2}(\frac{1}{y^2}-1)}(y^4-\frac{b^4}{a^4})^{-1}(1-\frac{b^4}{a^4})}}dy.
\end{equation}
Again, we need numerical analysis to obtain behavior of potential.

\begin{figure}[h!]
\begin{center}$
\begin{array}{cccc}
\includegraphics[width=100 mm]{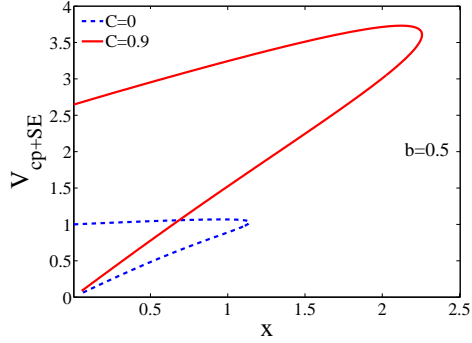}
\end{array}$
\end{center}
\caption{$V_{CP+SE}$ versus $x$ at finite temperature with $T_F\frac{R^2}{z_0}=z_0=1$.}
\label{fig:vxb}
\end{figure}

Fig. \ref{fig:vxb} shows the behavior of $V_{CP+SE}$ with respect to the distance in thermal case. It is clear that the distance increases with increasing potential and when deformation parameter contributes, the maximum value of the distance grows significantly. It is comparable with the Fig. \ref{fig:vxtzero} related to the zero temperature. For the given distance, the potential of the test particles grows under effect of deformation parameter, also the maximum value of $V_{CP+SE}$ increases. So, the nonzero value of $T$ or $c$ can lead to similar effects.\\
Again, according to the definition of $\alpha$ given by the equation (\ref{eq:alpha}) and with the following relation,
\begin{equation} \label{Ecr T}
E_{cr}(T)\mid_{c=0}=T_F\frac{R^2}{z_0^2}\sqrt{1-b^4},
\end{equation}
the total potential is given by  (\ref{eq:x t}) and (\ref{eq:v t}) as,
\begin{eqnarray}\label{Vtot T}
V_{tot}&=&V_{CP+SE}-Ex\nonumber\\
&=&2T_F\frac{R^2}{z_0}\int_{1}^{\frac{1}{a}}\frac{ae^{\frac{cz_0^2}{a^2}(\frac{1}{y^2}-\frac{1}{2})}(y^4-\frac{b^4}{a^4})-\frac{\alpha}{a}\sqrt{1-b^4}\sqrt{1-\frac{b^4}{a^4}}}{\sqrt{e^{\frac{cz_0^2}{a^2}(\frac{1}{y^2}-1)}(y^4-\frac{b^4}{a^4})^2-(y^4-\frac{b^4}{a^4})(1-\frac{b^4}{a^4})}}dy.
\end{eqnarray}

\begin{figure}[h!]
\begin{center}$
\begin{array}{cccc}
\includegraphics[width=100 mm]{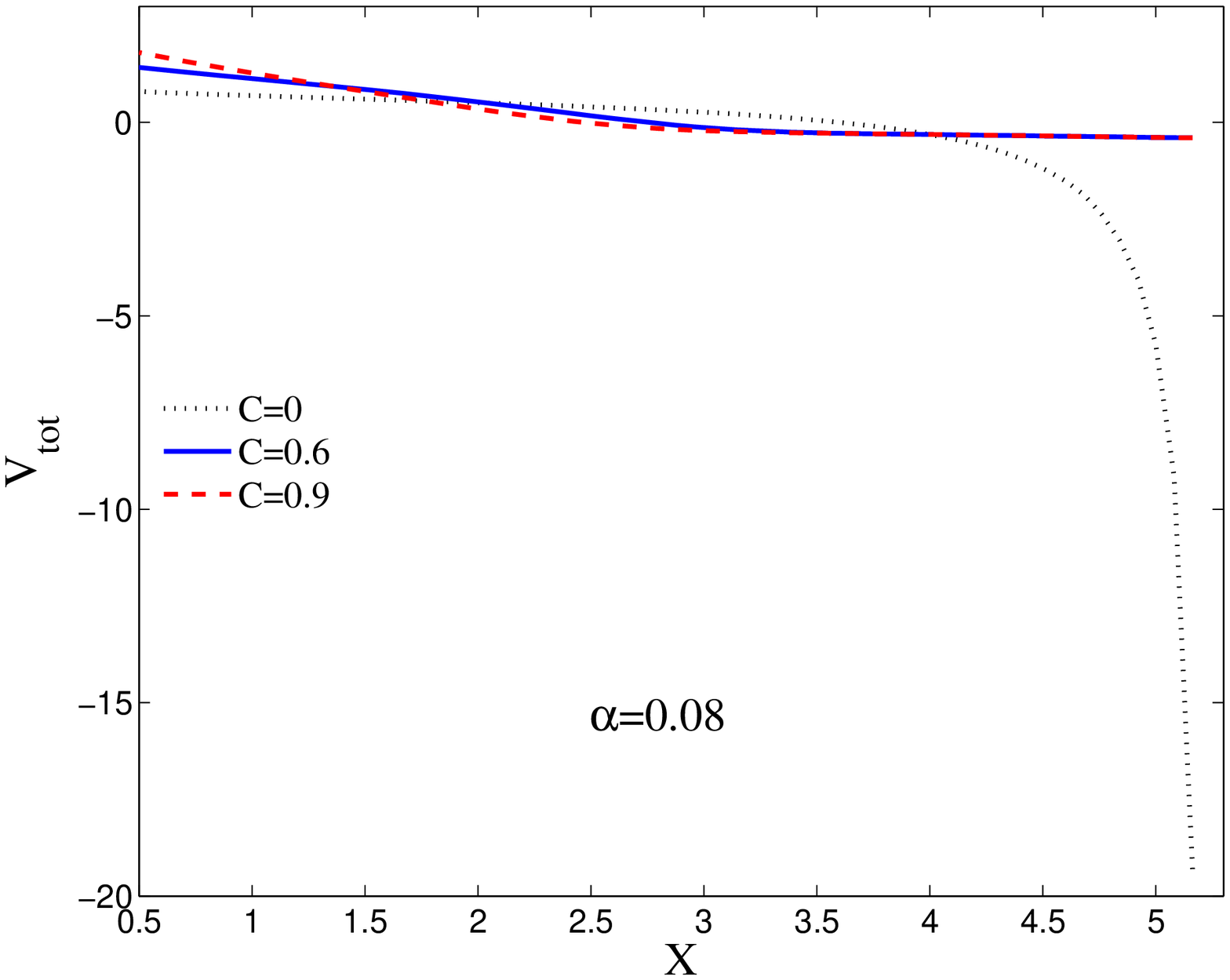}
\end{array}$
\end{center}
\caption{Total potential at finite temperature for $\alpha=0.08$ with $T_F\frac{R^2}{z_0}=z_0=1$ and $b=0.5$.}
\label{fig:vtot3-1}
\end{figure}

\begin{figure}[h!]
\begin{center}$
\begin{array}{cccc}
\includegraphics[width=100 mm]{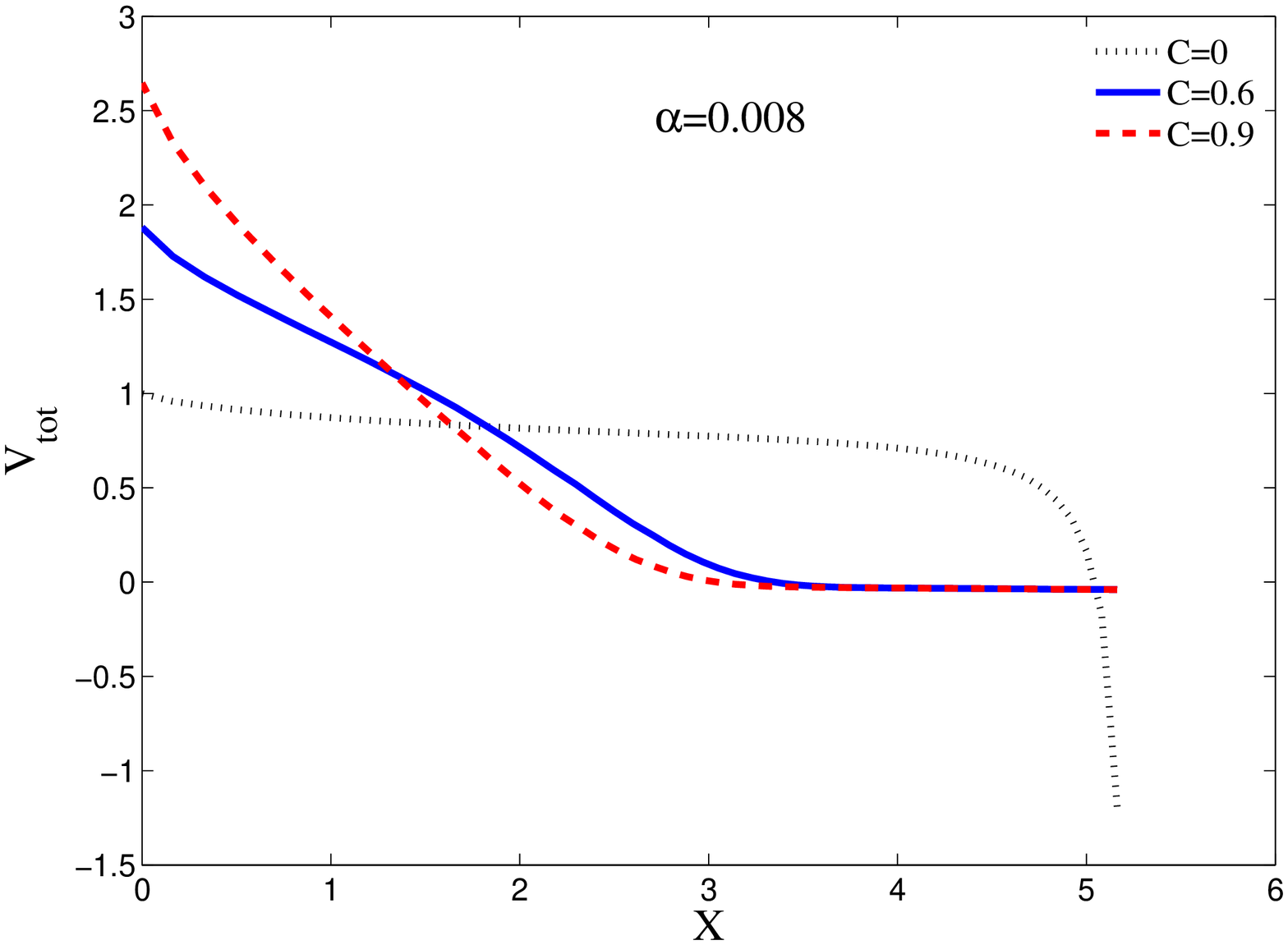}
\end{array}$
\end{center}
\caption{Total potential at finite temperature for $\alpha=0.008$ with $T_F\frac{R^2}{z_0}=z_0=1$ and $b=0.5$.}
\label{fig:vtot3-2}
\end{figure}

\begin{figure}[h!]
\begin{center}$
\begin{array}{cccc}
\includegraphics[width=100 mm]{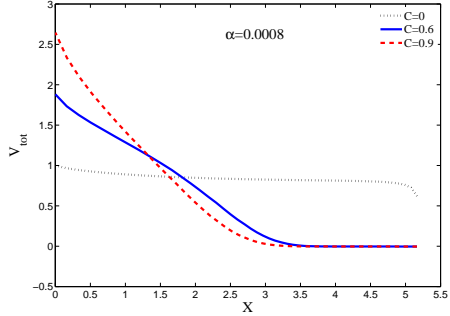}
\end{array}$
\end{center}
\caption{Total potential at finite temperature for $\alpha=0.0008$ with $T_F\frac{R^2}{z_0}=z_0=1$ and $b=0.5$.}
\label{fig:vtot3-3}
\end{figure}

\begin{figure}[h!]
\begin{center}$
\begin{array}{cccc}
\includegraphics[width=100 mm]{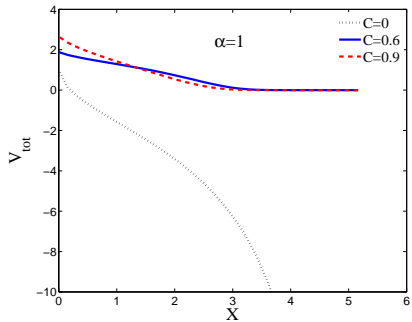}
\end{array}$
\end{center}
\caption{Total potential at finite temperature for $\alpha=1$ with $T_F\frac{R^2}{z_0}=z_0=1$ and $b=0.5$.}
\label{fig:vtot4}
\end{figure}

Figs. \ref{fig:vtot3-1},\ref{fig:vtot3-2},\ref{fig:vtot3-3} and Fig. \ref{fig:vtot4} are comparable with the Figs. \ref{fig:vtot1-1},\ref{fig:vtot1-2},\ref{fig:vtot1-3} and Fig. \ref{fig:vtot2} and show the effect of temperature. At the finite temperature the total potential does not have any potential barrier at deformed AdS. It is a strange result while the system with $c\neq 0$ is decaying for background electric field with any value. It can be interpreted that in a deformed AdS at a high enough temperature, any value of electric field can lead to a pair production with no potential barrier. In other words, there is a relation between $c$, $T$ and $E$. When both temperature and deformation parameter are nonzero and at least one of them has a large enough value, then any value of electric field can be interpreted as critical value of electric field.

\section{Pair production rate}
\begin{figure}[h!]
\begin{center}$
\begin{array}{cccc}
\includegraphics[width=100 mm]{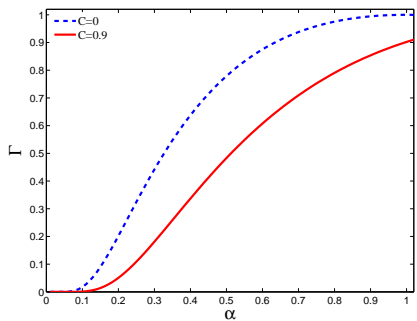}
\end{array}$
\end{center}
\caption{Pair production rate in terms of $\alpha$.}
\label{fig:11-1}
\end{figure}

\begin{figure}[h!]
\begin{center}$
\begin{array}{cccc}
\includegraphics[width=100 mm]{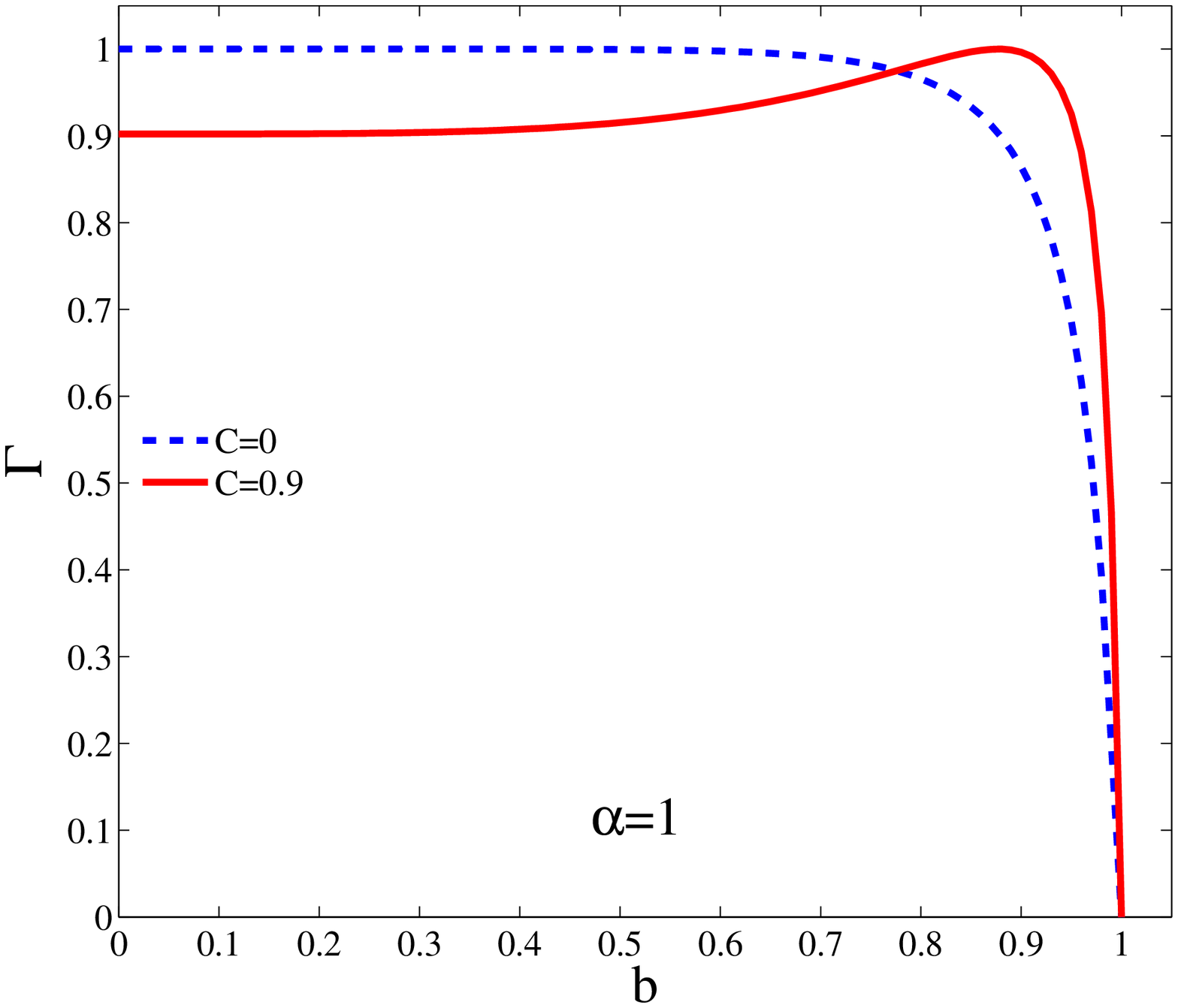}
\end{array}$
\end{center}
\caption{Pair production rate in terms of $b$ for $\alpha=1$.}
\label{fig:11-2}
\end{figure}

\begin{figure}[h!]
\begin{center}$
\begin{array}{cccc}
\includegraphics[width=100 mm]{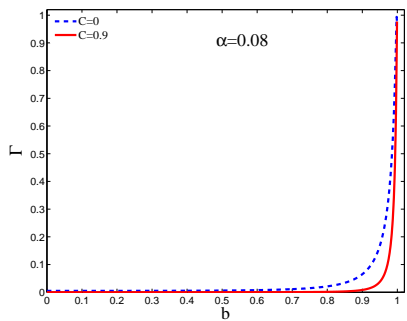}
\end{array}$
\end{center}
\caption{Pair production rate in terms of $\alpha=0.08$.}
\label{fig:11-3}
\end{figure}

Fig. \ref{fig:11-1} shows behavior of pair production rate with respect to $\alpha$. We show that increasing $\alpha$ corresponds to increasing electric field. It is clear that $\Gamma$ behaves exponentially with $\alpha$, but the intensity of pair production decreases with increasing deformation parameter. So, contribution of deformation parameter at some specified electric field, leads to delay in beginning of pair production process and decrease in pair production rate.
Figs. \ref{fig:11-2} and \ref{fig:11-3} show behavior of pair production rate against increasing temperature, for large and small values of electric field respectively. One can see the phenomenological behavior of them which are wholly different in two cases. For small values of $\alpha$ there is no serious difference between deformed AdS space and ordinary AdS space in pair production process. In both cases it depends on high temperature but at deformed AdS space the process confront more resistance. In the case of high enough temperature (near horizon), the pair production rate increased drammaticaly. This is comparable with large values of $\alpha$ or large values of electric field in which pair production rate is wholly considerable even in low temperature and again deformation parameter can interpreted as an opposed factor for pair production. One can see that there is a point of temperature thereafter, production rate falls. In AdS space it falls exponentially but when deformation parameter contributes, one can see an interesting extremity in pair production which thereafter it falls.

\section{Conclusion}
The purpose of the current study was to determine the effects of deformation parameter of AdS space on electrostatic potentials of a quark-antiquark pair in the Schwinger effect. Critical electric field which corresponds to the instability of vacuum also has been considered in the presence of deformation parameter.
We proceeded with the approach of \cite{pah}, where the analysis has been done by evaluating the classical action of a string solution attached to a probe D3-brane sitting at an intermediate position in the bulk. As the deformation parameter of the metric background has been contributed in the process and deformed the AdS space, remarkable changes to the results of the previous works with $c=0$ were found. We have performed numerical calculations in both zero and finite temperature cases to achieve the acceptable results.
In zero temperature case, we have seen the distance $x$ has maximum value when deformation parameter contributed. In addition, there is a degeneracy between the distance $x$ and the auxiliary parameter $a$ at nonzero value of $c$.
 Also, the $V_{CP+SE}$ against $x$ would have a maximum value in the presence of $c$, where a greater value of $c$, leads to a greater maximum of potential in a shorter range of distance.
 But in consideration of the total potential, we have seen reverse behavior of the pair production process in the presence of electric field or deformation parameter since to deform an AdS and to strength electric field have opposed effects.\\
At finite temperature, our results showed that critical electric field decreases with increasing temperature and increases with increasing deformation parameter. In addition, the critical electric field has a linear increase as a function of $\Delta m$, which is the difference between thermal and non-thermal mass of the string. So, with nonzero value of deformation parameter, the behavior of critical electric field as a function of $\Delta m$ has no change, except that functions with a greater value of $c$ changes with a stronger gradient.\\
Moreover, there are degeneracies between $x$ and $a$ at any value of deformation parameter (zero or nonzero), therefore nonzero values of $c$ or $T$ have similar results phenomenologically.  In spite of zero temperature in $c=0$ case, at finite temperature the distance $x$ has a maximum value which grows with $c$.  When we studied Coulomb potential against $x$, we saw graphs of total potentials with different values of  deformation parameter are coincident except at very small distances. Also, any value of electric field, can lead to a pair production with no potential barrier according to a relation between $c$, $T$ and $E$. In this manner, when both temperature and deformation parameter are nonzero and at least one of them has a large enough value, then any value of electric field can be interpreted as the critical value of electric field which leads to a catastrophic pair production thereafter.\\
All our above-mentioned results have been given in the presence of an external electric field. Thus, contribution of $c$ in the presence of magnetic field can be an interesting case of study, which left as future work.\\
 Finally, it may be interesting to study Schwinger effect and potential analysis for the hyperscaling violated background \cite{Hyper1,Hyper2,Hyper4,Hyper5,Hyper6}.\\\\

\textbf{Acknowledgement}\\
The authors are grateful very much to S. M. Rezaei for support and valuable activity in
numerical calculations.


\begin{thebibliography}{99}

\bibitem{jss}
J.~S.~Schwinger, {\it{On gauge invariance and vacuum polarization}}, Phys.\ Rev.\ D {\bf 82} (1951) 664.
\bibitem{wh}
W.~Heisenberg and H.~Euler, {\it{Consequences of Dirac's theory of positrons}}, Z.\ Phys.\ {\bf 98} (1936) 714
[\href{http://xxx.lanl.gov/abs/physics.hist-ph/0605038}
{{\tt arXiv:physics.hist-ph/0605038}}].
\bibitem{esf}
E.~S.~Fradkin and A.~A.~Tseytlin, {\it{Quantum string theory effective action }}, Nucl.\ Phys.\ B {\bf 261} (1985) 1.
\bibitem{cbm}
C.~Bachas and M.~Porrati, {\it{Pair creation of open string in an electric field}}, Phys.\ Lett.\ B {\bf 296} (1992) 77
[\href{http://xxx.lanl.gov/abs/hep-th/9209032}
{{\tt arXiv:hep-th/9209032}}].
\bibitem{P1}
J.~M.~Maldacena, {\it{The large N limit of superconformal field theories
and supergravity}}, Int.\ J.\ Theor.\ Phys.\ {\bf 38} (1999) 1113.
[\href{http://xxx.lanl.gov/abs/hep-th/9711200}
{{\tt arXiv:hep-th/9711200}}].
\bibitem{P2}
E.~Witten, {\it{Anti-de Sitter space and holography}}, Adv.\ Theor.\ Math.\
Phys.\ {\bf 2} (1998) 253.
[\href{http://xxx.lanl.gov/abs/hep-th/9802150}
{{\tt arXiv:hep-th/9802150}}].
\bibitem{P3}
S.~S.~Gubser, I.~R.~Klebanov, and A.~M.~Polyakov, {\it{Gauge theory correlators from noncritical string theory}}, Phys.\ Lett.\ B {\bf 428}
(1998) 105.
[\href{http://xxx.lanl.gov/abs/hep-th/9802109}
{{\tt arXiv:hep-th/9802109}}].
\bibitem{oas}
O.~Aharony, S.~S.~Gubster, J.~M.~Maldacena, H.~Ooguri and Y.~Oz, {\it{Large N Field Theories, String Theory and Gravity}} Phys.\ Rept {\bf 323}  (2000) 183
[\href{http://xxx.lanl.gov/abs/hep-th/9905111}
{{\tt arXiv:hep-th/9905111}}].
\bibitem{P4}
R.~A.~Janik, {\it{AdS/CFT and the dynamics of quark-gluon plasma}}, Prog.\ Theor.\ Phys.\ Suppl.\ {\bf 186} (2010) 534
[\href{http://xxx.lanl.gov/abs/1101.0419}
{{\tt arXiv:1101.0419}}].
\bibitem{P5}
J.~Sadeghi, et al., {\it{Application of AdS/CFT in Quark-Gluon Plasma}}, Advances in High Energy Physics
{\bf 2013} (2013) 759804
\bibitem{B1}
J.~Sadeghi and B.~Pourhassan, {\it{Drag force of moving quark at the
${\mathcal{N}} =2$ supergravity}}, JHEP {\bf0812} (2008) 026,
[\href{http://xxx.lanl.gov/abs/0809.2668}
{{\tt arXiv:0809.2668}}].
\bibitem{B2}
J.~Sadeghi,et al., {\it{Drag force of moving quark in STU background}}, Eur.\ Phys.\ J.\ C {\bf61} (2009)
527,
[\href{http://xxx.lanl.gov/abs/0901.0217}
{{\tt arXiv:0901.0217}}].
\bibitem{B3}
J.~Sadeghi, M.~R.~Setare, and B.~Pourhassan, {\it{Drag force with
different charges in STU background and AdS/CFT}}, J.\ Phys.\ G:\ Nucl.\
Part.\ Phys.\ {\bf36} (2009) 115005.
[\href{http://xxx.lanl.gov/abs/0905.1466}
{{\tt arXiv:0905.1466}}].
\bibitem{B4}
K.~Bitaghsir Fadafan, B.~Pourhassan and J.~Sadeghi, {\it{Calculating the
jet-quenching parameter in STU background}}, Eur.\ Phys.\ J.\ C {\bf 71}
(2011) 1785,
[\href{http://xxx.lanl.gov/abs/1005.1368}
{{\tt arXiv:1005.1368}}].
\bibitem{AQ1}
A.~Karch, E.~Katz, D.~T.~Son, M.~A.~Stephanov, {\it{Linear Confinement and AdS/QCD}}, Phys.\ Rev.\ D {\bf74} (2006) 015005
[\href{http://xxx.lanl.gov/abs/hep-ph/0602229}
{{\tt arXiv:hep-ph/0602229}}].
\bibitem{AQ2}
S.~J.~Brodsky, G.~F.~de Teramond, {\it{AdS/CFT and QCD}}, slac-pub-12361,
[\href{http://xxx.lanl.gov/abs/hep-th/0702205}
{{\tt arXiv:hep-th/0702205}}].
\bibitem{AQ3}
C.~Csaki, M.~Reece, J.~Terning, {\it{The AdS/QCD Correspondence: Still Undelivered}},  JHEP {\bf0905} (2009)  067
[\href{http://xxx.lanl.gov/abs/0811.3001}
{{\tt arXiv:0811.3001}}].
\bibitem{AQ4}
A.~V.~Ramallo, {\it{Introduction to the AdS/CFT correspondence}},
[\href{http://xxx.lanl.gov/abs/1310.4319}
{{\tt arXiv:1310.4319}}].
\bibitem{B5}
B.~Pourhassan, J.~Sadeghi, {\it{STU/QCD Correspondence}}, Canadian Journal of Physics, {\bf 91} (2013) 995
[\href{http://xxx.lanl.gov/abs/1205.4254}
{{\tt arXiv:1205.4254}}].
\bibitem{stp}
A.~S.~Gorsoy, K.~A.~Saraikin and K.~G.~Selivanov, {\it{Schwinger type processes via brane and their gravity duals}}, Nucl.\ Phys.\ B {\bf 628} (2002) 270 [arxiv: 0110178 [hep-th]].
[\href{http://xxx.lanl.gov/abs/hep-th/0110178}
{{\tt arXiv:hep-th/0110178}}].
\bibitem{rhw}
J.~Ambjorn and Y.~Makeenko, {\it{Remarks on holographic Wilson loops and the Schwinger effect }}, Phys.\ Rev.\ D {\bf 85} (2012) 061901  [arxiv:1112.5606 [hep-th]].
[\href{http://xxx.lanl.gov/abs/1112.5606}
{{\tt arXiv:1112.5606}}].
\bibitem{coce}
S.~Bolognesi, F.~Kiefer and E.~Rabinovici, {\it{Comments on critical electric and
magnetic fields from holography}}, JHEP {\bf 01} (2013) 174
[\href{http://xxx.lanl.gov/abs/1210.4170}
{{\tt arXiv:1210.4170}}].
\bibitem{hds}
Y.~Sato and K.~Yoshida, {\it{Holographic description of the Schwinger effect in electric
and magnetic fields }}, JHEP {\bf 04} (2013) 111
[\href{http://xxx.lanl.gov/abs/1303.0112}
{{\tt arXiv:1303.0112}}].
\bibitem{secp}
Y.~Sato and K.~Yoshida, {\it{Holographic Schwinger effect in confining phase }}, JHEP {\bf 09} (2013) 134
[\href{http://xxx.lanl.gov/abs/1306.5512}
{{\tt arXiv:1306.5512}}].
\bibitem{uah}
Y.~Sato and K.~Yoshida, {\it{Universal aspects of holographic Schwinger effect in general backgrounds }}, JHEP {\bf 12} (2013) 051
[\href{http://xxx.lanl.gov/abs/1309.4629}
{{\tt arXiv:1309.4629}}].
\bibitem{ppr}
D.~Kawai, Y.~Sato and K.~Yoshida, {\it{The Schwinger pair production rate in confining
theories via holography }}, Phys.\ Rev.\ D {\bf 89} (2014) 101901
[\href{http://xxx.lanl.gov/abs/1312.4341}
{{\tt arXiv:1312.4341}}].
\bibitem{hdse}
D.~Kawai, Y.~Sato, and K.~Yoshida, {\it{A holographic description of the Schwinger effect in a confining gauge theory}}, Int.\ J.\ Mod.\ Phys.\ A {\bf 30} (2015) 1530026
[\href{http://xxx.lanl.gov/abs/1504.00459}
{{\tt arXiv:1504.00459}}].
\bibitem{npp}
M.~Sakaguchi, H.~Shin and K.~Yoshida, {\it{No pair production of open strings in a
plane-wave background}}, Phys.\ Rev.\ D {\bf 90} (2014) 066009
[\href{http://xxx.lanl.gov/abs/1402.2048}
{{\tt arXiv:1402.2048}}].
\bibitem{hsed}
W.~Fischler, P.~H.~Nguyen, J.~F.~Pedraza, and W.~Tangarife, {\it{Holographic Schwinger effect in de Sitter space}}, Phys.\ Rev.\ D {\bf 91} (2015) 086015
[\href{http://xxx.lanl.gov/abs/1411.1787}
{{\tt arXiv:1411.1787}}].
\bibitem{nre}
K.~Bitaghsir Fadafan and F.~Saiedi, {\it{On Holographic Non-relativistic Schwinger Effect}}, Eur.\ Phys.\ J.\ C {\bf 75} (2015) 612
[\href{http://xxx.lanl.gov/abs/1504.02432}
{{\tt arXiv:1504.02432}}].
\bibitem{scen}
M.~Ghodrati, {\it{Schwinger Effect and Entanglement Entropy in Confining Geometries}}, Phys.\ Rev.\ D {\bf 92} (2015) 065015
[\href{http://xxx.lanl.gov/abs/1506.08557}
{{\tt arXiv:1506.08557}}].
\bibitem{vie}
K.~Hashimoto and T.~Oka, {\it{Vacuum instability in electric fields via AdS/CFT:
Euler-Heisenberg lagrangian and Planckian thermalization }}, JHEP, {\bf 10} (2013) 116
[\href{http://xxx.lanl.gov/abs/1307.7423}
{{\tt arXiv:1307.7423}}].
\bibitem{pah}
Y.~Sato and K~Yoshida, {\it{Potential analysis in holographic Schwinger effect}}, JHEP {\bf 08} (2013) 002
[\href{http://xxx.lanl.gov/abs/1304.7917}
{{\tt arXiv:1304.7917}}].
\bibitem{hse}
G.~W.~Semenoff and K.~Zarembo, {\it{Holographic Schwinger effect}}, Phys.\ Rev.\ Lett.\ {\bf 107} (2011) 171601
[\href{http://xxx.lanl.gov/abs/1109.2920}
{{\tt arXiv:1109.2920 }}].
\bibitem{CLSS}
C.~Ewerz, L.~Lin, A.~Samberg, K.~Schade, {\it{Holography for Heavy Quarks and Mesons at Finite Chemical Potential}}, PoS (CPOD2014) 037
[\href{http://xxx.lanl.gov/abs/1511.04006}
{{\tt arXiv:1511.04006}}].
\bibitem{Ex1}
D.~-f.~Zeng, {\it{Heavy Quark Potentials in Some Renormalization Group Revised AdS/QCD Models}}, Phys.\ Rev.\ D {\bf 78} (2008) 126006
[\href{http://xxx.lanl.gov/abs/0805.2733}
{{\tt arXiv:0805.2733}}].
\bibitem{Ex2}
H.~J.~Pirner, B.~Galow, {\it{Equivalence of the AdS-Metric and the QCD Running Coupling}},  Phys.\ Lett.\ B {\bf679} (2009) 51
[\href{http://xxx.lanl.gov/abs/0903.2701}
{{\tt arXiv:0903.2701}}].
\bibitem{cor}
O.~Andreev,  {\it{$\frac{1}{q^2}$ Corrections and Gauge/String Duality}}, Phys.\ Rev.\ D {\bf 73} (2006) 107901
[\href{http://xxx.lanl.gov/abs/hep-th/060317}
{{\tt arXiv:060317 [hep-th]}}].
\bibitem{hqp}
O.~Andreev and V.~I.~Zakharov, {\it{Heavy-Quark Potentials and AdS/QCD}}, Phys.\ Rev.\ D {\bf 74} (2006) 025023
[\href{http://xxx.lanl.gov/abs/hep-ph/0604204}
{{\tt arXiv:0604204 [hep-ph]}}].
\bibitem{Xing}
X.~Wu, {\it{Notes on holographic Schwinger effect}}, JHEP {\bf1509} (2015) 044
[\href{http://xxx.lanl.gov/abs/1507.03208}
{{\tt arXiv:1507.03208}}].
\bibitem{GI}
K.~Ghoroku, M.~Ishihara, {\it{Holographic Schwinger Effect and Chiral condensate in SYM Theory}}, JHEP {\bf1609} (2016) 011
[\href{http://xxx.lanl.gov/abs/1604.05025}
{{\tt arXiv:1604.05025}}].
\bibitem{hsc}
J.~Polchinski and M.~Strassler, {\it{Hard Scattering and Gauge/String Duality }}, Phys.\ Rev.\ Lett {\bf 88} (2002) 031601
[\href{http://xxx.lanl.gov/abs/hep-th/0109174}
{{\tt arXiv:0109174 [hep-th]}}].
\bibitem{scv}
O.~Andreev, {\it{Scaling Violation and Gauge/String Duality}}, Phys.\ Rev.\ D {\bf 73} (2006) 046010
[\href{http://xxx.lanl.gov/abs/hep-th/0512282}
{{\tt arXiv:0512282 [hep-th]}}].
\bibitem{tsst}
O.~Andreev and V.~I.~Zakharov, {\it{The Spatial String Tension, Thermal Phase Transition, and AdS/QCD}}, Phys.\ Lett.\ B {\bf 645} (2007) 437
 [\href{http://xxx.lanl.gov/abs/hep-ph/0607026}
{{\tt arXiv:0607026 [hep-ph]}}].
\bibitem{tec}
J.~Sadeghi and S.~Tahery, {\it{The effects of deformation parameter on thermal width of moving quarkonia in plasma}}, JHEP {\bf 06}  (2015)  204
[\href{http://xxx.lanl.gov/abs/1412.8332}
{{\tt arXiv:1412.8332}}]



\bibitem{AAM}
I.~K.~Affleck, O.~Alvarez and N.~S.~Manton, {\it{Pair Production At Strong Coupling In
Weak External Fields}} Nucl.\ Phys.\ B {\bf197} (1982) 509.



\bibitem{msh}
S.~J.~Rey and J.~T.~Yee, {\it{Macroscopic strings as heavy quarks in large N gauge theory
and anti-de Sitter supergravity}}, Eur.\ Phys.\ J.\ C {\bf 22} (2001) 379
[\href{http://xxx.lanl.gov/abs/hep-th/9803001}
{{\tt arXiv:9803001 [hep-th]}}].
\bibitem{wll}
J.~M.~Maldacena, {\it{Wilson loops in large N field theories}}, Phys.\ Rev.\ Lett {\bf 80} (1998) 4859
[\href{http://xxx.lanl.gov/abs/hep-th/9803002}
{{\tt arXiv:9803002 [hep-th]}}].





\bibitem{wpl}
S.~J.~Rey, S.~Theisen, J.~T.~Yee , {\it{Wilson-Polyakov Loop at Finite Temperature in Large N Gauge Theory and Anti-de Sitter Supergravity }}, Nucl.\ Phys.\ B {\bf 527} (1998) 171
[\href{http://xxx.lanl.gov/abs/hep-th/9803135}
{{\tt arXiv:9803135 [hep-th]}}].
\bibitem{eih}
K.~Hashimoto, T.~Oka and A.~Sonoda, {\it{Electromagnetic instability in holographic QCD }}, JHEP {\bf 06} (2015) 001
[\href{http://xxx.lanl.gov/abs/1412.4254}
{{\tt arXiv:1412.4254}}].


\bibitem{Hyper1}
S.~Kachru, X.~Liu, M.~Mulligan, {\it{Gravity dual of Lifshitz-like fixed points}}, Phys.\ Rev.\ D {\bf 78} (2008) 106005
[\href{http://xxx.lanl.gov/abs/0808.1725}
{{\tt arXiv:0808.1725}}].
\bibitem{Hyper2}
J.~Sadeghi, B.~Pourhassan, F.~Pourasadollah, {\it{Thermodynamics of Schr\"{o}dinger black holes with hyperscaling violation}}, Phys.\ Lett.\ B {\bf 720} (2013) 244
[\href{http://xxx.lanl.gov/abs/1209.1874}
{{\tt arXiv:1209.1874}}].
\bibitem{Hyper4}
J.~Sadeghi, B.~Pourhassan, A.~Asadi, {\it{Application of hyperscaling violation in QCD}}, Can.\ J.\ Phys.\ {\bf 92} (2014) 280
\bibitem{Hyper5}
M.~Cadoni, S.~Mignemi,  {\it{Phase transition and hyperscaling violationfor scalar black branes}} JHEP {\bf1206}  (2012) 056
[\href{http://xxx.lanl.gov/abs/1205.0412}
{{\tt arXiv:1205.0412}}].
\bibitem{Hyper6}
J.~Sadeghi, B.~Pourhassan, A.~Asadi, {\it{Thermodynamics of string black hole with hyperscaling violation}}, Eur.\ Phys.\ J.\ C 74 (2014) 2680
[\href{http://xxx.lanl.gov/abs/1209.1235}
{{\tt arXiv:1209.1235}}].
\end{thebibliography}
\end{document}